\documentstyle[aps]{revtex}
\begin{document}
\draft

\title{Comment on ``Correlation between Compact Radio Quasars and
Ultrahigh Energy Cosmic Rays''}
\author{C.\ M.\ Hoffman}
\address{Physics Division\\
Los Alamos National Laboratory\\
Los Alamos, NM 87545}
\date{\today}
\maketitle

\begin{abstract}
\end{abstract}

In the paper ``Correlation between Compact Radio Quasars and Ultrahigh
Energy Cosmic
Rays,'' \cite{1} Farrar and Biermann argue that there is a strong
correlation between the
direction of the five highest-energy cosmic-ray events and compact,
radio-loud quasars.
Because this result, if true, would have profound implications, it has been
widely reported
in the popular scientific press \cite{2}. This Comment shows that the
analysis in Ref.
\cite{1} contains several inconsistencies and errors so that the
significance of any such
correlation is certainly greatly overestimated and perhaps nonexistent.

A proton, nucleus or photon above 10$^{20}$ eV is exceedingly unlikely to
travel further
than 50 Mpc due to interactions with the cosmic microwave background: this
is the
Greisen-Zatsepin-Kuzmin limit \cite{3}. The compact, radio-loud quasars
(CQSOs) in
question have red shifts ranging from $0.29 < z < 2.18$.  Thus new physics
is required if
the observed high-energy cosmic rays did originate in CQSOs.  The authors
suggest two
possibilities:

\begin{enumerate}
\item The existence of a new neutral long-lived hadron with a mass of a few
GeV,
such as a light gluino \cite{4}, and
\item Neutrinos with a mass of a few eV interacting resonantly with
dark-matter
neutrinos clustered in the halo of either our galaxy or the local cluster of
galaxies producing a $Z$-boson.  This results in an ultrahigh-energy hadron or
photon that, in turn, produces the observed extensive air shower \cite{5}.
\end{enumerate}
In both of these scenarios, the high-energy cosmic ray should point back to
its source.  It is
this hypothesis that is tested in Ref. \cite{1}, with the conclusion that
the five high-energy
cosmic-ray events examined are all aligned with CQSOs, and that the
probability that these
alignments are coincidental is 0.005.

The authors chose to examine all high-energy cosmic-ray events whose energy
is at
least 1$\sigma$ above $8 \times 10^{19}$ eV and whose direction is known
with a solid
angle resolution, $\Delta\Omega$, of 10 deg$^{2}$ or better.  While only
qualitative
reasons are given for these specific criteria, it is striking that the
energy cut barely excludes
two events from Haverah Park with energies of $(1.02 \pm 0.3) \times
10^{20}$ eV and
$(1.05 \pm 0.3) \times 10^{20}$ eV, and one from Yakutsk with an energy of
$(1.1 \pm
0.4) \times 10^{20}$ eV \cite{6}.  Remarkably, one of the events used in
the analysis
of Ref. \cite{1} also fails the cut: this is the event labeled Ag110, which
has a measured
energy of $(1.10 \times 10^{20})$ eV $\pm$ 30\%.

It is unclear what Farrar and Biermann mean by ``solid angle resolution.''
In Table I of
Ref. \cite{1}, the $\Delta\Omega$s given for two events (Ag210 and Ag110)
contain the
true event directions with 68\% probability, while the $\Delta\Omega$s
given for the other
three events (FE320, HP120, HP105) contain the true event direction with
only $\sim
12$\% probability.  In addition, the values of $\Delta\Omega$ in Table I of
Ref. \cite{1}
for FE320 and HP120 are incorrectly calculated.

These discrepancies call into question whether the events used in further
analysis have
been selected in an fair, unbiased manner, and whether the subsequent
analysis, in which
the probability that randomly distributed objects with a given surface
density would appear
aligned with the five events is calculated, is correct.

Even if the analysis of Ref. \cite{1} is correct, the statistical
significance given for the
alignment (0.005) is not.  The authors formulate the hypothesis of correlations 
between the cosmic-ray directions and CQSOs (as opposed to other possible 
astrophysical objects) because a correlation with a CQSO had already been noted 
for the 320 TeV Fly's Eye event (FE320) \cite{7}.  
An event used to formulate a hypothesis may not be used
to test that
hypothesis.  Eliminating FE320 from the analysis lowers the statistical
significance of the
alignment to 0.03.  It is this number, not 0.005, that correctly assesses
the evidence for the
hypothesis that the high-energy cosmic-ray events point back to compact
quasars,
assuming that the selection criteria are unbiased and that the rest of the
analysis in Ref.
\cite{1} has been done correctly.

I gratefully acknowledge illuminating conversations with C. Sinnis.   This
work was
partially supported by the U.S.\ Department of Energy, Los Alamos National
Laboratory,
and the University of California.


\end{document}